\documentclass[pra,twocolumn,superscriptaddress,longbibliography]{revtex4-2}

\usepackage{amsmath,amssymb,braket,graphicx,siunitx,booktabs}
\usepackage{multirow}
\usepackage[dvipsnames]{xcolor}
\usepackage[colorlinks=true,citecolor=blue,linkcolor=blue,urlcolor=blue,backref=true]{hyperref}
\usepackage{oplotsymbl}
\usepackage{pifont}

\newcommand{\mb}{\mathbf}
\newcommand{\intf}[3]{\int_{#1}^{#2}\text{d}#3\ }
\newcommand{\proj}[2]{\left|#1\right>\left<#2\right| }
\newcommand{\rme}{\mathrm{e}}
\newcommand{\abs}[1]{\left|#1\right|}

\DeclareFontFamily{U}{calligra}{}
\DeclareFontShape{U}{calligra}{m}{n}{<->callig15}{}

\begin{document}

\title{Light-Harvesting Efficiency Cannot Depend on Optical Coherence in the Absence of Orientational Order}

\author{Dominic M. Rouse}
\affiliation{School of Physics and Astronomy, University of Glasgow, Glasgow G12 8QQ, United Kingdom}
\affiliation{Department of Physics and Astronomy, University of Manchester, Manchester M13 9PL, United Kingdom}

\author{Adesh Kushwaha}
\affiliation{School of Chemistry and University of Sydney Nano Institute, University of Sydney NSW 2006, Australia}

\author{Stefano Tomasi}
\affiliation{School of Chemistry and University of Sydney Nano Institute, University of Sydney NSW 2006, Australia}

\author{Brendon W. Lovett}
\affiliation{SUPA, School of Physics and Astronomy, University of St Andrews, St Andrews KY16 9SS, United Kingdom}

\author{Erik M. Gauger}
\affiliation{SUPA, Institute of Photonics and Quantum Sciences, Heriot-Watt University, Edinburgh EH14 4AS, United Kingdom}

\author{Ivan Kassal}
\email[]{ivan.kassal@sydney.edu.au}
\affiliation{School of Chemistry and University of Sydney Nano Institute, University of Sydney NSW 2006, Australia}


\begin{abstract}

The coherence of light has been proposed as a quantum-mechanical control for enhancing light-harvesting efficiency. In particular, optical coherence can be manipulated by changing either the polarization state or spectral phase of the light. Here, we show that, in weak light, light-harvesting efficiency cannot be controlled using any form of optical coherence in molecular light-harvesting systems and, more broadly, those comprising orientationally disordered sub-units and operating on longer-than-ultrafast timescales. Under those conditions, optical coherence does not affect light-harvesting efficiency, meaning that it cannot be used for control. Specifically, polarization-state control is lost in disordered samples or when the molecules reorient on the timescales of the light-harvesting, and spectral-phase control is lost when the efficiency is time-averaged for longer than the optical coherence time. In practice, efficiency is always averaged over long times, meaning that coherent optical control is only possible through polarisation in systems with orientational order.

\end{abstract}

\maketitle

Controlling the efficiency of light-harvesting processes using various manifestations of quantum-mechanical coherence has been an active field of research because of the promise of using engineered quantum systems to improve the efficiency of solar energy conversion \cite{trebbia2022tailoring,creatore2013efficient,dorfman2013photosynthetic,fruchtman2016photocell,rouse2019optimal,wertnik2018optimizing,gelbwaser2017thermodynamic,fang2022coherent,policht2022hidden,romero2014quantum,romero2017quantum,cao2020quantum,wertnik2018optimizing,tscherbul2018non,tomasi2019coherent,tomasi2021environmentally,svidzinsky2011enhancing}. Some proposals use excitonic coherence---the coherence within the reduced density matrix of the light-harvesting molecules~\cite{tomasi2020classification}---to enhance the efficiency of excitation capture or retention \cite{creatore2013efficient,dorfman2013photosynthetic,fruchtman2016photocell,rouse2019optimal,wertnik2018optimizing,trebbia2022tailoring,davidson2020dark,davidson2022eliminating,svidzinsky2011enhancing}. Other methods instead propose altering the optical coherence---the coherence within the incident light---which can be achieved by manipulating the exciting field \cite{manvcal2010exciton,tomasi2019coherent,tomasi2021environmentally}. It is the latter form with which we are concerned here. 

Optical coherent control is the manipulation of system observables---such as the light-harvesting efficiency---through changes to the coherence properties of the field \cite{shapiro2012quantum}. The focus on coherence rules out trivial types of optical control that are based on changing the total power or the spectrum of the light.
The variables that affect the coherence but not the spectrum can be seen in the Fourier expansion of a general electric field, 
\begin{equation}\label{eq:Etrans}
\mb{E}(t)=\frac{1}{\sqrt{2\pi}}\intf{-\infty}{\infty}{\omega}
\tilde{\mb{E}}(\omega)\rme^{-i\omega t},
\end{equation}
where the frequency components are \begin{equation}\label{eq:Ew}
    \tilde{\mb{E}}(\omega)=\rme^{i\tilde{\phi}(\omega)}\tilde{E}_0(\omega)\tilde{\mb{n}}(\omega).
\end{equation} 
In each frequency component, $\tilde{\phi}(\omega)$ is the spectral phase, while $\tilde{\mb{n}}(\omega)$ indicates the polarization direction. $\tilde{\mb{n}}(\omega)$ is a complex unit vector to allow non-linear polarizations. To ensure $\tilde{\phi}(\omega)$ is uniquely defined, we assume the $x$ component $\tilde{n}_x(\omega)$ is real. Eq.~\eqref{eq:Etrans} is restricted to a single point in space to avoid integrals over propagation directions; as we discuss below, this is exact in the electric dipole approximation, which is almost always appropriate for light-harvesting.

In general, optical fields are stochastic, meaning that each of the variables $\tilde{E}_0(\omega)$, $\tilde{\phi}(\omega)$ and $\tilde{\mb{n}}(\omega)$ is, in each realisation of the optical ensemble, a random variable drawn from some distribution.
The two strategies for optical coherent control are changes to the distributions from which $\tilde{\phi}(\omega)$ and $\tilde{\mb{n}}(\omega)$ (but not $\tilde{E}_0(\omega)$) are drawn, because they leave the power spectrum $\tilde{P}(\omega)=\langle \tilde{\mb{E}}(\omega)\cdot\tilde{\mb{E}}^*(\omega)\rangle$ unaffected, where $\langle\cdot\rangle$ is the ensemble average over the stochastic realizations of the field. We call these two approaches spectral phase control and polarization control.

Both spectral phase control (through $\tilde{\phi}(\omega)$) and polarization control (through $\tilde{\mb{n}}(\omega)$) are possible in certain circumstances.
For example, spectral phase control has been used to modify energy flow in both photosynthetic~\cite{Herek2002, Wohlleben2005} and artificial~\cite{Savolainen2008,Kuroda2009} light harvesters, as well as to control the isomerization of retinal~\cite{retinal}.
However, all of these examples rely on multi-photon processes achieved in high-intensity laser experiments~\cite{Lavigne2017}.

By contrast, practical light-harvesting takes place in weak light, where spectral phase control is also referred to as one-photon phase control (OPPC) \cite{spanner2010communication,pachon2013coherent,pachon2013mechanisms,lavigne2019considerations,pachon2013incoherent,arango2013communication,manvcal2010exciton,mukamel2013coherent,am2014scaling,shapiro1995quantum,wu2008quantum}. Weak light means that a semi-classical first-order perturbation theory in the light-matter interaction is accurate and, therefore, that at most one excitation exists in a molecule at a time. The weak-field approximation is exceptionally accurate for light-harvesting because of the low intensity of sunlight; for example, typical chlorophyll excitation rates vary from $10^{-4}\,\mathrm{s^{-1}}$ in overcast conditions to $10\,\mathrm{s^{-1}}$ in bright sunlight \cite{van2017limits}. In closed systems, OPPC is possible for observables that do not commute with the light-independent part of the Hamiltonian \cite{spanner2010communication,pachon2013coherent}. 
Open quantum systems additionally allow for OPPC if the bath interaction couples the population dynamics to the excitonic coherences \cite{spanner2010communication,pachon2013coherent,pachon2013mechanisms}.

Polarization control is also possible in general. A simple example would be a system with all transition dipole moments aligned; light polarized in the direction of the dipoles would be absorbed, while light with a perpendicular polarization would not. 
A less trivial possibility is control through only the degree of coherence between different polarization directions: light-harvesting efficiency in a dimer can be doubled by using polarized light compared to unpolarized light of the same intensity \cite{tomasi2019coherent,tomasi2021environmentally}. These efficiency gains can be reinforced by stronger coupling to the vibrational baths \cite{tomasi2021environmentally}, conditions relevant for many candidate light-harvesting systems \cite{fruchtman2016photocell}. 

However, the control examples given above all assume special circumstances that are unlikely to apply in practical light-harvesting. In particular, OPPC proposals rely on the ability to carry out measurements on ultrafast timescales, while the polarization control in refs.~\cite{tomasi2019coherent,tomasi2021environmentally} assumes a definite relative orientation between the light-harvesting molecules and the incoming light. By contrast, light-harvesting almost always occurs over long periods of time in disordered systems; for example, biological chromophores are randomly aligned with respect to sunlight and their efficiency is averaged over the lifetime of the organism, whether days or years. 

\begin{figure}[tb]
    \centering
    \includegraphics[width=\columnwidth]{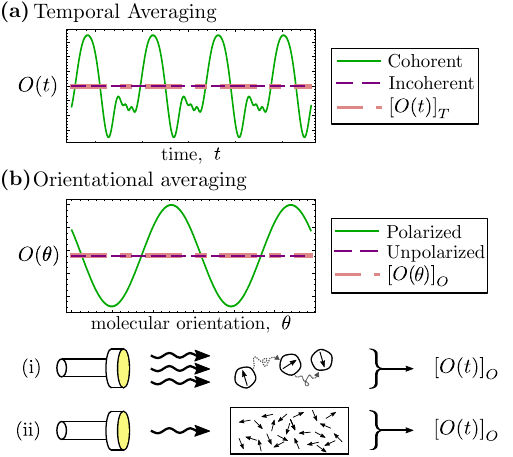}
    \caption{\textbf{Loss of optical coherent control upon averaging.}
    \textbf{(a)}~Temporal averaging: the expectation value of an observable is plotted over time for a light field with different spectral coherences. $[O(t)]_T$ is the temporal average of either `Coh.' or `Incoh.'.
    \textbf{(b)}~Orientational averaging: an observable is plotted as a function of molecular orientation for different polarizations of light. $[ O(\theta)]_O$ is the orientational average of either polarization states. Control through the polarization state is lost whether \textbf{(i)} a series of measurements are taken on a reorienting molecule or \textbf{(ii)} a single measurement is taken of a disordered sample.}
    \label{fig:Schematic}
\end{figure}

Here, we show that in realistic light-harvesting systems, neither the efficiency nor any other observable can be controlled by either type of optical coherent control. In realistic cases, the temporal averaging rules out spectral phase control, while the remaining polarization control is rendered impossible by the orientational averaging. These averages are illustrated in Fig.~\ref{fig:Schematic}. We also show what types of coherent control are possible if only one type of averaging takes place and we show the possibility of optical coherent control---both spectral and polarization---when the averages are partial.

\textit{Model.} We consider the Hamiltonian
\begin{equation}\label{eq:H}
H(t)=H_S+H_B+H_{SB}+H_O(t),
\end{equation}
where $H_S$ describes a light-harvesting sub-unit, $H_B$ describes the environmental bath, which is coupled to the sub-unit through $H_{SB}$ and causes dephasing and dissipation among the excited states of the sub-unit but has no influence on its ground state. The sub-unit could be, for example, a molecule tumbling in solution, or a single unit of a disordered structure, as illustrated in Fig.~\ref{fig:Schematic}(b). For light-harvesting in weak light, we can describe the light-matter interaction using the semi-classical Hamiltonian
\begin{equation}\label{eq:HO}
H_O(t)=-\sum_\mu\mb{d}_\mu\cdot\left(\mb{E}(t)\proj{\mu}{G}+\mb{E}^*(t)\proj{G}{\mu}\right),
\end{equation}
where $G$ and $\mu$ are the ground and excited eigenstates of $H_S$, $\mb{d}_\mu$ is the transition dipole moment from $G$ to $\mu$, and $\mb{E}(t)$ is the electric field. Assuming that the wavelength of light is larger than the sub-unit, we can ignore spatial field dependence, i.e., we make the electric dipole approximation and describe the field using Eq.~\eqref{eq:Etrans}. Because of the weakness of the field, we can restrict our consideration to the zero- and single-excitation manifolds. Lastly, we have ignored permanent dipole moment interactions in Eq.~\eqref{eq:HO} because these do not induce transitions to leading order \cite{may2023charge}. 

Under Eq.~\eqref{eq:H}, the total density operator of the system and bath evolves as
\begin{equation}\label{eq:psi}
\frac{\text{d}}{\text{d}t}\rho(t)=\mathcal{L}_{SB}\rho(t)+\sum_{\mu,i}\left( \mb{d}_{\mu}\cdot\mb{e}_i\right)E_i(t)\mathcal{V}_\mu\rho(t)+\text{H.c.},
\end{equation}
where $\mathcal{L}_{SB}\rho(t)=-i[H_S+H_B+H_{SB},\rho(t)]$ is the combined system-bath Liouvillian superoperator, $\mathcal{V}_\mu\rho(t)=-i[\proj{\mu}{G},\rho(t)]$ is the optical interaction superoperator, and `H.c.' denotes the Hermitian conjugate \cite{manvcal2010exciton}. We have decomposed the field $\mb{E}(t)$ into its Cartesian components $E_{i}(t)$ along the directions of the unit vectors $\mb{e}_i$ for $i\in\{x,y,z\}$.

In Supporting Information Section~S1, we show that integrating Eq.~\eqref{eq:psi} to second order in the light-matter coupling $H_O$ and performing the ensemble average yields
\begin{align}\label{eq:rho2}
\left\langle\rho(t)\right\rangle=\rho(t_0)+\intf{t_0}{\infty}{\tau}&\intf{t_0}{\infty}{\tau'}\sum_{\mu,\nu}R_{\mu\nu}(\tau,\tau')\nonumber
\\&\times \Lambda_{\mu\nu}(t-\tau,\tau-\tau')\rho(t_0),
\end{align}
where $\rho(t_0)$ is the initial density operator, which we take to be a product of the electronic ground state with a thermal bath state. The term that is first-order in $H_O$ is absent because, for optical fields, $\langle E(t)\rangle=0$. The optical coherence matrix $R_{\mu\nu}(\tau,\tau')$, and the propagation matrix $\Lambda_{\mu\nu}(s,s')$, describe two aspects of the light-matter interaction. On the one hand,
\begin{equation}\label{eq:Aij}
R_{\mu\nu}(\tau,\tau')= \sum_{i,j}\kappa_{\mu\nu ij}G_{ij}(\tau,\tau')
\end{equation}
describes the coherence properties of the light and its interaction with the system, with 
\begin{equation}\label{eq:O}
	\kappa_{\mu\nu ij}=\left(\mb{d}_\mu\cdot\mb{e}_i\right)\left(\mb{d}_\nu\cdot\mb{e}_j\right)
\end{equation}
being an orientation factor and
\begin{equation}\label{eq:G}
G_{ij}(\tau,\tau')=\langle E_i(\tau)E_{j}^*(\tau')\rangle
\end{equation}
being the two-time correlation function of the field.
On the other hand, the propagation matrix 
\begin{equation}\label{eq:wtL}
    \Lambda_{\mu\nu}(s,s')=\mathcal{G}_{SB}(s)\mathcal{V}_\mu\mathcal{G}_{SB}(s')\mathcal{V}_\nu
\end{equation}
describes the full system-bath evolution using the Green's function $\mathcal{G}_{SB}(t)=\Theta(t)\exp(\mathcal{L}_{SB}t)$, with $\Theta(t)$ being the Heaviside step function. Although Eq.~\eqref{eq:rho2} is perturbative in $H_O$, the influence of the bath on the time evolution is incorporated exactly through the propagation matrix~\cite{manvcal2010exciton}.

Since optical coherence is best parameterised in Fourier space, we insert the Fourier transforms of the fields in Eq.~\eqref{eq:Etrans} into Eq.~\eqref{eq:rho2}, yielding (see Supporting Information Section~S2)
\begin{multline}\label{eq:rhoO}
	\left\langle\rho(t)\right\rangle=\rho(t_0)+\intf{-\infty}{\infty}{\omega_1}	\intf{-\infty}{\infty}{\omega_2}\rme^{-i\omega_1 t}\\
	\times\sum_{\mu,\nu}\tilde{R}_{\mu\nu}(\omega_1-\omega_2,\omega_2)\tilde{\Lambda}_{\mu\nu}(\omega_1,\omega_2)\rho(t_0).
\end{multline}
Functions with tildes and frequency arguments relate to the same functions with temporal arguments by Fourier transforms, using the convention in Eq.~\eqref{eq:Etrans}. Importantly, the optical coherent control parameters $\kappa_{\mu\nu ij}$ and $G_{ij}(\tau,\tau')$ influence the density operator only through the optical coherence matrix $\tilde{R}_{\mu\nu}(\omega,\omega')$ and not through the propagation matrix $\tilde{\Lambda}_{\mu\nu}(\omega,\omega')$ (see Eqs.~\eqref{eq:Aij}, \eqref{eq:wtL}). Therefore, the central question is how $\tilde{R}_{\mu\nu}(\omega,\omega')$ is affected by the orientational and temporal averages.

\textit{Orientational averaging.} 
Orientational averaging occurs if the molecules do not have a fixed spatial orientation, which can occur in two ways, illustrated in Fig.~\ref{fig:Schematic}(b). First, the light-harvesting sub-unit could be tumbling randomly between different light-absorption events, as, for example, if the system consists of molecules in solution. Second, a system could be composed of static sub-units oriented randomly, such as the molecules in a disordered organic semiconductor. Orientational averaging is the ensemble average over the possible orientations of the tumbling sub-unit, or of the static absorbers in the disordered sample.

Mathematically, orientational averaging involves averaging the orientation of the transition dipole moments $\mb{d_\mu}$ within the optical coherence matrix $\tilde{R}_{\mu\nu}(\omega,\omega')$ over all angles, which only affects the orientation factor $\kappa_{\mu\nu ij}$. To distinguish orientational and temporal averages from the optical ensemble average, we denote them with square brackets with subscript `$O$' or `$T$'. The quantity currently of interest is $[\kappa_{\mu\nu ij}]_O$.

If the sub-unit is tumbling or disordered in three dimensions, $\mb{d_\mu}$ and $\mb{d_\nu}$ take on all possible values such that $\mb{d_\mu}\cdot \mb{d_\nu}$ is fixed. Then every contribution $(\mb{d_\mu}\cdot \mb{e}_i)(\mb{d_\nu}\cdot \mb{e}_{j})$ to the orientational average in $[\kappa_{\mu\nu ij}]_O$ where $i \ne j$ is cancelled by a contribution of $(\mb{d_\mu} \cdot \mb{e}_i)((-\mb{d_\nu}) \cdot \mb{e}_{j})$ from another configuration formed by rotating the dipoles by \ang{180} around $\mb{e}_i$. The average of the surviving terms is
\begin{align}
    \left[\kappa_{\mu \nu ij}\right]_O &= \left[ \left(\mb{d}_\mu\cdot\mb{e}_i\right)\left(\mb{d}_\nu\cdot\mb{e}_i\right)   \right]_O \delta_{ij} \nonumber\\ 
    & = \tfrac13 \left(\mb{d}_\mu\cdot\mb{d}_\nu\right)\delta_{ij}, \label{eq:kavg}
\end{align}
where the last equality follows from 
$
    \mb{d_\mu}\cdot \mb{d_\nu}=\sum_{i \in \{x,y,z\}}(\mb{d_\mu}\cdot \mb{e}_i)(\mb{d_\nu}\cdot \mb{e}_i),
$
where each of the three terms on the right-hand side must be equal under orientational averaging because none of the three directions is preferred. Therefore, orientational averaging removes the terms with $i\neq j$ from the optical coherence matrix.

\textit{Temporal averaging.} The quantities of interest for systems operating over long timescales are most commonly time-averaged observables. We now prove that when time-independent observables are averaged over times longer than the coherence time of the incident light, spectral coherence control becomes impossible. 

The time-averaged expectation value of a time-independent operator $O$ is $[O]_T=[\text{Tr}(O\langle\rho(t)\rangle)]_T=\text{Tr}(O[\langle\rho(t)\rangle]_T)$, where the time-averaged density operator is
\begin{equation}
    \label{eq:rhob}\left[\langle\rho(t)\right\rangle]_T=\frac{1}{T}\intf{t_c-\frac{T}{2}}{t_c+\frac{T}{2}}{t}\left\langle\rho(t)\right\rangle,
\end{equation}
with the averaging duration $T$, centered at $t_c$. The only time-dependent factor in Eq.~\eqref{eq:rhoO} is $\rme^{-i\omega_1t}$,  which implies that the integral over time in Eq.~\eqref{eq:rhob} is
\begin{align}
	\frac{1}{T}\intf{t_c-\frac{T}{2}}{t_c+\frac{T}{2}}{t}\rme^{-i\omega_1 t}&=\rme^{-i\omega_1t_c} \operatorname{sinc}\left(\frac{\omega_1 T}{2}\right)\nonumber\\
    &\approx \rme^{-i\omega_1 t_c}\frac{2\pi}{T}\delta(\omega_1),
\label{eq:DD}
\end{align}
where the second line holds for $\omega_1T\gg 1$. Since $\omega_1$ is integrated over in Eq.~\eqref{eq:rhoO}, the averaging time must be large enough to satisfy $\omega_1T\gg 1$ for all values of $\omega_1$ for which $|\tilde{G}(\omega_1-\omega_2,\omega_2)|$ is non-negligible. This implies the requirement that $T\gg \tau_\text{coh}$, the coherence time of the light, in agreement with earlier results~\cite{lavigne2019considerations}. For sunlight, $\tau_\text{coh}\sim 1~\text{fs}$ \cite{ricketti2022coherence}, and for lasers $\tau_\text{coh}$ can be up to $10~\text{s}$ \cite{oelker2019demonstration}, both of which are much shorter than the days or years over which the performance of light-harvesting systems is typically evaluated.

Inserting Eq.~\eqref{eq:DD} into Eq.~\eqref{eq:rhoO} causes the terms with unequal frequencies to vanish. Therefore, temporal averaging affects the optical coherence matrix as
\begin{equation}
    \label{eq:RT}\left[\tilde{G}_{ij}(\omega,\omega')\right]_T=\tilde{G}_{ij}(\omega,\omega) \delta(\omega-\omega').
\end{equation}

\textit{Effects of averaging on the coherent controls.}
We can now assess the effects of orientational and temporal averaging---described by Eqs.~\eqref{eq:kavg} and \eqref{eq:RT}---on the two types of optical coherent control: through the polarization state $\tilde{\mb{n}}(\omega)$, and the spectral phase $\tilde{\phi}(\omega)$.
These results are summarised in Table~\ref{tab:avg}.

Most importantly, all forms of optical coherent control are impossible under simultaneous orientational and temporal averaging.
Substituting the general electric-field expression from Eq.~\eqref{eq:Etrans} into Eq.~\eqref{eq:Aij} yields
\begin{multline}\label{eq:R2}
    \tilde{R}_{\mu\nu}(\omega,\omega')=\tilde{E}_0(\omega)\tilde{E}^*_0(\omega')\rme^{i\left(\tilde{\phi}(\omega)-\tilde{\phi}(\omega')\right)}
    \\
    \times\sum_{i,j}\kappa_{\mu\nu ij}\tilde{n}_i(\omega)\tilde{n}_j^*(\omega').
\end{multline}
Using Eqs.~\eqref{eq:kavg} and \eqref{eq:RT} as well as the normalisation $\sum_{i}|\tilde{n}_i(\omega)|^2=1$, under simultaneous orientational and temporal averaging Eq.~\eqref{eq:R2} becomes
\begin{equation}\label{eq:ROT}
    \left[\tilde{R}_{\mu\nu}(\omega,\omega')\right]_{O,T}=\tfrac13 |\tilde{E}_0(\omega)|^2\left(\mb{d}_\mu\cdot\mb{d}_\nu\right)\delta(\omega-\omega'),
\end{equation}
which no longer depends on $\tilde{\mb{n}}(\omega)$ or $\tilde{\phi}(\omega)$.
This scenario---the complete loss of optical coherent control---is by far the most relevant to practical light-harvesting. All biological light-harvesting systems and most existing artificial ones involve averages over many randomly oriented sub-units and over timescales of days, much longer than the coherence time of any practical light source.

We can also consider the effects of each type of averaging in isolation. Under only temporal averaging,
\begin{equation}
    \left[\tilde{R}_{\mu\nu}(\omega,\omega')\right]_T=|\tilde{E}_0(\omega)|^2\sum_{i,j}\kappa_{\mu\nu ij}\tilde{n}_i(\omega)\tilde{n}_j^*(\omega)\delta(\omega-\omega'),
\end{equation}
which is independent of $\tilde{\phi}(\omega)$, indicating that under temporal averaging, spectral-phase control is impossible. However, temporal averaging does not remove control through the polarization state.

By contrast, under only orientational averaging,
\begin{multline}\label{eq:RO2}
    \left[\tilde{R}_{\mu\nu}(\omega,\omega')\right]_O=\tfrac13 \tilde{E}_0(\omega)\tilde{E}_0^*(\omega')\left(\mb{d}_\mu\cdot\mb{d}_\nu\right)\rme^{i\left(\tilde{\phi}(\omega)-\tilde{\phi}(\omega')\right)}\\
    \times \sum_i\tilde{n}_i(\omega)\tilde{n}_i^*(\omega').
\end{multline}
Here, both types of control are still possible. However, it is very common for the polarization state to be frequency-independent, in which case orientational averaging removes polarization control, because $\sum_i\tilde{n}_i(\omega)\tilde{n}_i^*(\omega') = |\tilde{\mb{n}}|^2 = 1$ in Eq.~\eqref{eq:RO2}.

\begin{table}[tb]
	\centering
    \renewcommand{\arraystretch}{1.35}
	\begin{tabular*}{5.8cm}{@{\extracolsep{\fill}} l c c } 
			\toprule
    Average & $\tilde{\phi}(\omega)$ control & $\tilde{\mb{n}}(\omega)$ control \\
    \midrule 
   $\left[\left\langle\rho(t)\right\rangle\right]_O$ & $\squad$ & $\squadfillha$ \\ 
   $\left[\left\langle\rho(t)\right\rangle\right]_T$ & $\squadfill$ & $\squad$  \\
   $\left[\left\langle\rho(t)\right\rangle\right]_{O,T}$ & $\squadfill$ & $\squadfill$   \\
			\bottomrule
		\end{tabular*}
		\caption{\textbf{Effects of complete orientational and temporal averaging on the possibility of optical coherent control.} $\squadfill$ denotes that the average always prevents the control. $\squad$ denotes that the control remains possible under the average. $\squadfillha$ denotes that the average only prevents the control if the control is frequency-independent.}
		\label{tab:avg}
\end{table}

\textit{Optical coherent control with partial averaging.} 
So far, we have shown that complete averaging causes a loss of coherent control; we now show that both types of optical coherent control are possible if the corresponding averaging is partial. Of course, the partial-averaging results recover the loss of control summarised in Table~\ref{tab:avg} in the limit of completely averaged orientations and long-time averaging. Throughout this section, we use the light-harvesting efficiency as the observable of interest, although our results hold for any time-independent and isotropic observable of the system.

\begin{figure}[tb]
	\centering
	\includegraphics[width=\columnwidth]{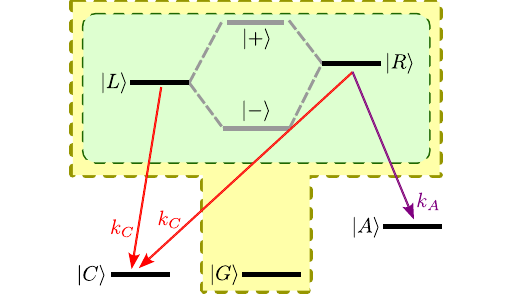}
	\caption{\textbf{Model used for numerical simulations.} The yellow area indicates levels affected by $H_O$ (including the eigenstates of $H_S$ in grey) and the green area indicates levels affected by $H_{SB}$. The arrows show the transitions caused by the Lindblad operators.}
	\label{fig:model}
\end{figure}

As a concrete example of our general findings, we demonstrate partial averaging in a model system that is a frequently studied one in the context of light-harvesting \cite{creatore2013efficient,dorfman2013photosynthetic,fruchtman2016photocell,rouse2019optimal,tomasi2019coherent,tomasi2021environmentally}, and is shown in Fig.~\ref{fig:model}. We consider a rigid molecule, approximated as a dimer of two-level sites with excited states left $\ket{L}$ and right $\ket{R}$. The system starts in the ground state $\ket{G}$ and is illuminated by a pulse. Each site has a transition dipole moment $\mb{d}_a$ for $a\in\{L,R\}$ and the total Hamiltonian is given in Eq.~\eqref{eq:H}. 

The Hamiltonian describing the molecule is
\begin{equation}
  H_S=\sum_{a\in\{L,R\}}\epsilon_a\proj{a}{a}+\Omega\left(\proj{L}{R}+\proj{R}{L}\right),
\end{equation}
where $\epsilon_a$ are the site energies and the excitonic coupling $\Omega$ is constant due to the rigidity of the molecule. In the eigenbasis, $H_S=\sum_{\mu\in\{+,-\}}\epsilon_\mu\left|\mu\right\rangle\left\langle 
\mu\right|$, where the eigenstates $\ket{\pm}$ have energies $\epsilon_\pm=(\epsilon_L+\epsilon_R\pm\tilde{\epsilon})/2$ with $\tilde{\epsilon}=((\epsilon_R-\epsilon_L)^2+4\Omega^2)^{1/2}$. 

We assume each dipole has an independent but identical bath, typically a vibrational one, described by
\begin{equation}
    H_B=\sum_{k}\sum_{a\in\{L,R\}}\omega_{k} b^\dagger_{ka} b_{ka},
\end{equation}
where $b_{ka}$ annihilates a bath excitation of mode $k$ on site $a$. The system-bath interaction is in the site basis,
\begin{equation}
    H_{SB}=\sum_{k}\sum_{a\in\{L,R\}}\proj{a}{a} g_{k}\left( b^\dagger_{ka} + b_{ka}\right),
\end{equation}
and, in the continuum limit, is described by the spectral density $J(\omega)=\sum_{k}g_{k}^2\delta(\omega-\omega_{k})$, for which we choose a super-Ohmic form,
\begin{equation}
    J(\omega)=\frac{S}{\omega_{c}^2}\omega^3\rme^{-\omega/\omega_{c}},
\end{equation}
where $\omega_{c}$ is the cut-off frequency and $S$ is the Huang-Rhys parameter. The super-Ohmic form models the broad low-frequency contributions from molecule vibrations \cite{renger2002relation}. In the numerical analysis, we choose typical values of the parameters in the Hamiltonian, inspired by photosynthetic molecules \cite{higgins2017quantum}, but the qualitative conclusions we draw regarding partial averaging are independent of these. We solve the bath evolution using Redfield theory, which is justified for our spectral density choice and weak system-bath interactions \cite{breuer2002theory}.

To calculate the light-harvesting efficiency, we introduce two additional levels which are initially unpopulated: a collector state $\ket{C}$ and an acceptor state $\ket{A}$. Once created in the molecule, an excitation will eventually transfer to either the collector or acceptor state via incoherent processes. Transfer to the acceptor models a successful light-harvesting event that contributes to the efficiency, while transfer to the collector models the decay of the excitation, i.e., failed light-harvesting. For simplicity, we add these processes through Lindblad operators, 
\begin{equation}
L(a,b)=k_b\left(\ket{b}\bra{a}\rho\ket{a}\bra{b}-\tfrac{1}{2}\left\{\ket{a}\bra{a},\rho\right\}\right),
\end{equation}
which models an incoherent transition from $\ket{a}$ to $\ket{b}$ at rate $k_b$, with $\{\cdot,\cdot\}$ the anti-commutator. In our theory, Lindblad operators are included within the evolution terms $\mathcal{G}_{SB}(t)$ in Eq.~\eqref{eq:wtL}. The specific Lindblad processes we add are shown in Fig.~\ref{fig:model}: decay from $\ket{L}$ and $\ket{R}$ to $\ket{C}$ at rate $k_C$ and transfer from $\ket{R}$ to $\ket{A}$ at rate $k_A$. That is, we include $L(L,C)$, $L(R,C)$, $L(R,A)$.

The efficiency of the light-harvesting process can be defined in several closely related ways. Here, we consider a quantum efficiency, one proportional to the probability the excitation makes it to the acceptor state instead of decaying. For simplicity, we choose
\begin{equation}\label{eq:eff}
    \eta(t)= \left\langle\rho_{AA}(t)\right\rangle,
\end{equation}
where $\rho_{aa}(t)=\braket{a|\rho(t)|a}$. This definition is not unique; for example, one could define efficiency as $\eta'(t)=\langle\rho_{AA}(t)\rangle / (\langle\rho_{AA}(t)\rangle+\langle\rho_{CC}(t)\rangle)$, which agrees with Eq.~\eqref{eq:eff} if one excitation is created in the system. However, the definition in Eq.~\eqref{eq:eff} is easier to measure, more useful in practice, and an observable, unlike $\eta'(t)$, which is a ratio of two observables. The latter difference leads to a subtlety in that some polarization control of $\eta'(t)$ can be lost under partial orientational averages even though observables remain controllable (see Supporting Information Section~S3).

\begin{figure}[tb]
	\centering
	\includegraphics[width=\columnwidth]{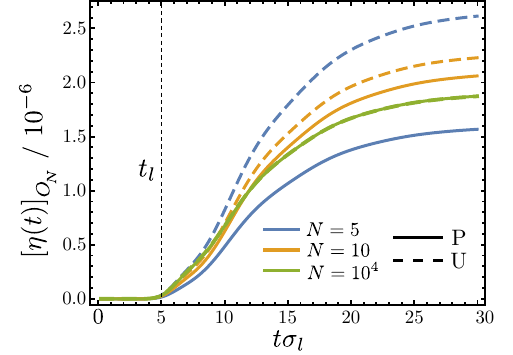}
	\caption{\textbf{Polarization control is possible for partial orientational averaging.}
	The efficiency for polarized (P) and unpolarized (U) light in three dimensions after averaging over $N$ possible dimer orientations. For small $N$, one can control the efficiency by changing the polarization state, but the control is lost at large $N$. $[\eta(t)]_{O_N}$ is the average over $N$ iterations, such that $[\eta(t)]_{O_\infty}=[\eta(t)]_O$. Parameters used: $\epsilon_L=1$~eV, $\epsilon_R=1.2$~eV, $\Omega=0.05$~eV, $\sigma_l=\tilde{\epsilon}=\sqrt{(\epsilon_R-\epsilon_L)^2+4\Omega^2}$, $S=0.5$, $\omega_{c}=0.5\tilde{\epsilon}$, $k_t=k_r=0.1\tilde{\epsilon}$, $\xi=\pi/3$.}
	\label{fig:pol}
\end{figure}

We can now show that polarization control of the efficiency is possible under partial orientational averaging. For concreteness, we choose a frequency-independent and real $\tilde{\mb{n}}$, corresponding to linearly polarised light. We do not consider spectral phase in this example, setting $\tilde{\phi}(\omega)=0$. As the field envelope, we choose a transform-limited Gaussian pulse of frequency width $\sigma_l$, central time $t_l$, and central frequency $\omega_l$,
\begin{equation}\label{eq:E}
    E(t)=\rme^{-i\omega_l\left(t-t_l\right)}\exp\left(-\tfrac{1}{2}\sigma_l^2\left(t-t_l\right)^2\right).
\end{equation}

We consider two examples of partial orientational averaging.
First, we assume that the molecule is free to tumble in three dimensions, showing that as the number of orientations averaged over becomes large, polarization control is lost. Second, we impose greater orientational restriction by considering a molecule confined to rotations within a plane, a situation that preserves some control through the polarization even when all of the permitted orientations are averaged.

\begin{figure}[tb]
	\centering
	\includegraphics[width=\columnwidth]{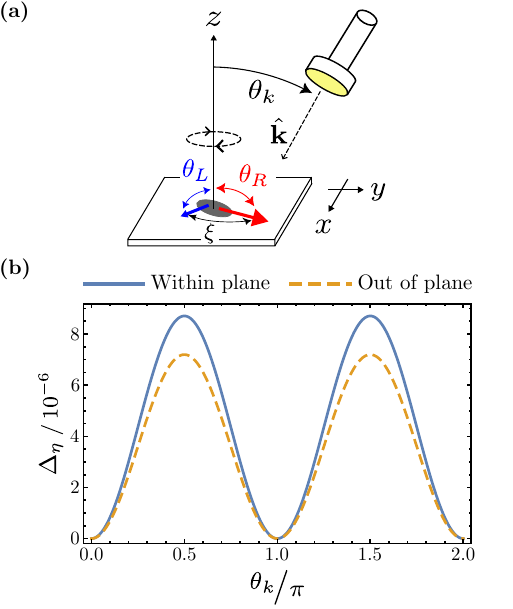}
	\caption{\textbf{Polarization control of the efficiency is possible in two dimensions.} (a) A schematic showing the relevant angles, and that the molecule can only rotate about the $z$-axis. (b) The available polarization control $\Delta_{\eta}$ (defined in Eq.~\eqref{eq:PC}) over the light-harvesting efficiency as a function of the propagation direction of the light, $\theta_k$, averaged over all possible orientations $(N\to\infty)$ in two dimensions. In-plane dipoles have $\theta_L=\theta_R=\pi/2$ and out-of-plane have $\theta_L=0.6\pi$ and $\theta_R=0.45\pi$. In both cases, $\xi=\pi/3$ and $t_f=30/\sigma_l$ are fixed. Other parameters are as in Fig.~\ref{fig:pol}.}
	\label{fig:pol2d}
\end{figure}

Fig.~\ref{fig:pol} shows the efficiency for molecules free to tumble in three dimensions. We choose linearly polarized light propagating in the $z$-direction, meaning that the polarization is in the $xy$ plane, $\tilde{\mb{n}}=\{\tilde{n}_x,(1-\tilde{n}_x^2)^{1/2},0\}$, determined by the single control parameter $\tilde{n}_x$. We consider both $x$-polarized light with $\tilde{n}_x=1$ and unpolarized light, represented as $\tilde{n}_x=\cos(\chi)$ where $\chi$ is drawn uniformly at random from $[0,2\pi]$. We average the efficiency over $N$ random orientations of the molecule, which keep the relative angle between the site dipoles fixed at $\xi=\pi/3$. Fig.~\ref{fig:pol} shows that for few iterations $N$, the efficiencies for the unpolarized and polarized light are different, but, in agreement with Table~\ref{tab:avg}, become equal at all times when $N$ is large. Thus, polarization control in three dimensions is possible under partial orientational averaging. 

Fig.~\ref{fig:pol2d} shows the efficiency when the molecules are restricted to reorient within a plane (the $xy$ plane, see Fig.~\ref{fig:pol2d}(a)). We again compare $x$-polarized and unpolarized light; however, because the model is no longer symmetric with respect to the propagation direction of the light, we perform the calculations with the light propagating at each angle $\theta_k$ in the $yz$ plane. For each $\theta_k$, we calculate the available optical coherent control averaged over all possible orientations ($N\to\infty$), defined as
\begin{equation}\label{eq:PC}
        \Delta_{\eta}=\left|\left[\eta^p(t_f)\right]_O-\left[\eta^u(t_f)\right]_O\right|,
\end{equation}
where we use the time $t_f=30\sigma_l^{-1}$ and $\eta^{p,u}(t)$ are the efficiencies for the polarized and unpolarized fields. A larger $\Delta_\eta$ indicates more available polarization control over the efficiency. 

Fig.~\ref{fig:pol2d}(b) shows that polarization control of the efficiency is only impossible if the light is propagating perpendicular to the plane. Otherwise, the amount of polarization control varies with propagation direction, and is maximal when the light is parallel to the plane. This further shows that the limits in Table~\ref{tab:avg} are only guaranteed for complete---i.e., three-dimensional and across all orientations---averages. In Supporting Information Section~S3, we further show that, even though $\eta(t)$ is controllable, the control of $\eta'(t)$ can nevertheless be lost.

\begin{figure}[tb]
    \centering
    \includegraphics[width=0.48\textwidth]{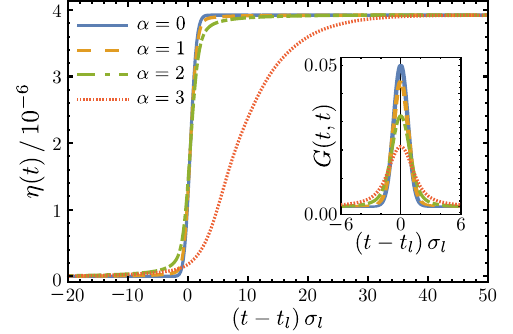}
    \caption{\textbf{Spectral phase control of the light-harvesting efficiency $\eta(t)$ is possible at short times.} The spectral coherence of the light is varied by changing $\alpha$, which alters $\eta(t)$ at short times. However, $\eta(t)$ is constant at long times, regardless of $\alpha$. Parameters are the same as in Fig.~\ref{fig:pol}, except that the dipoles are orthogonal and within the $xy$ plane, and the pulse is centered at $t_l=100\sigma_l$. $t_l$ is large so that as little light power as possible is cut off before numerical integration starts at $t_0$. However, because $t_0-t_l$ cannot be infinite in numerical calculations, we make tiny scaling corrections to $G(\omega,\omega)$ to maintain a fixed power (these are purely numerical artefacts). The inset shows the pulse shape intensity, which broadens from a transform-limited Gaussian (spectrally coherent, $\alpha=0$) to Lorentzian (spectrally incoherent, $\alpha\gg1$).}
    \label{fig:temp}
\end{figure}

Finally, we show that spectral phase control is possible for transient times, but is lost under long-time averaging. For this example, we assume that the light is unidirectional and linearly polarized, so that $\tilde{\phi}(\omega)$ is the only remaining control. The effect of the spectral phase is contained within the spectral correlation function
\begin{equation}\label{eq:Gphi}
    \tilde{G}(\omega,\omega')=\tilde{E}_0(\omega) \tilde{E}^*_0(\omega')\left\langle \rme^{i\Delta\tilde{\phi}(\omega,\omega')}\right\rangle,
\end{equation}
where $\Delta\tilde{\phi}(\omega,\omega')=\tilde{\phi}(\omega)-\tilde{\phi}(\omega')$ and $\tilde{E}_0(\omega)$ is a Gaussian envelope with frequency width $\sigma_l$. For concreteness, we consider a field with
\begin{equation}\label{eq:Gwwp}
\tilde{G}(\omega,\omega')=\tilde{E}_0(\omega) \tilde{E}^*_0(\omega')\rme^{-\frac{1}{2}\alpha|\omega-\omega'|},
\end{equation}
where $\alpha$ quantifies the spectral coherence.   

For numerical purposes, we construct the spectral phase functions $\tilde{\phi}(\omega)$ that results in Eq.~\eqref{eq:Gwwp} on a grid with spacing $\delta\omega$. Starting from initial point $\omega_0$ with phase $\tilde{\phi}(\omega_0)$, remaining phases are built using $\tilde{\phi}(\omega_{n\pm 1})=\tilde{\phi}(\omega_n)+\delta \tilde{\phi}$ where the phase increments $\delta \tilde{\phi}$ are sampled from a normal distribution of mean 0 and variance $\alpha\,\delta\omega$. Because $\Delta\tilde{\phi}(\omega_{n\pm M},\omega_n)$ is a sum of $M$ independent and identically distributed Gaussian random variables, its variance is $M\alpha\,\delta\omega=\alpha\abs{\omega_{n\pm M}-\omega_n}$.
Since the characteristic function of a Gaussian random variable $x$ of mean 0 and variance $\Sigma^2$ is
$\langle \rme^{i x t}\rangle=\rme^{-\frac{1}{2}\Sigma^2 t^2},$
we find $\langle \rme^{i\Delta\tilde{\phi}(\omega,\omega')}\rangle = \rme^{-\frac{1}{2}\alpha|\omega-\omega'|}$, giving Eq.~\eqref{eq:Gwwp}.

Fig.~\ref{fig:temp} shows the transient light-harvesting efficiency for several values of $\alpha$, showing that $\eta(t)$ depends on---and can be controlled by---the spectral coherence of the light. However, at long times, $\eta(t)$ is independent of $\alpha$. For our choice of electric field, the coherence time $\tau_\text{coh}$ is approximately $\sigma_l^{-1}$ when $\alpha=0$ and grows with $\alpha$. As shown in Fig.~\ref{fig:temp}, spectral phase control is possible within the coherence time of the laser, but drops off rapidly when $t-t_l\gg \sigma_l^{-1}$ in agreement with the discussion following Eq.~\eqref{eq:DD}. Since light-harvesting systems operate on the timescale of days and years, it is clear that for practical purposes spectral phase control is impossible.

\textit{Conclusion.} Optical coherence could be used to affect light-harvesting in two ways, through the polarization or the spectral phase of the light.
Our results show that both of these mechanisms become impossible when the observable of interest---especially the light-harvesting efficiency---is averaged simultaneously over molecular orientations and time. This situation is by far the most common in natural and artificial light-harvesting systems. However, this does not mean that optical coherence can never affect light-harvesting efficiency. If the system is only subject to one type of averaging, Table~\ref{tab:avg} summarises which types of control are possible. Similarly, if one performs a partial orientational or temporal average---including if molecules are restricted to two dimensions---then some degree of optical coherent control remains possible. 

This paper provides a new perspective on the long-standing debate regarding the role that quantum coherence plays in light-harvesting. In particular, we have shown that optical coherence is irrelevant to the efficiency of processes in naturally occurring complexes and most practical devices. By contrast, as other works have shown~\cite{creatore2013efficient,dorfman2013photosynthetic,fruchtman2016photocell,rouse2019optimal,wertnik2018optimizing,trebbia2022tailoring,davidson2020dark,davidson2022eliminating,svidzinsky2011enhancing}, excitonic coherence---whether in the site or energy basis---offers more promising avenues for optimising light-harvesting performance.

\section*{Supporting Information}
Section~S1: Derivation of the time-domain density operator. Section~S2: Fourier-space density operator. Section~S3: Controlling the efficiency defined as the ratio of two observables.

\section*{Acknowledgments} 
D.M.R.~was supported by EPSRC Grant EP/T517896. A.K.~and I.K.~were supported by the Australian Research Council (DP220103584), a Sydney Quantum Academy scholarship, and by computational resources from the National Computational Infrastructure (Gadi). E.M.G.~acknowledges support from the EPSRC under Grant No. EP/T007214/1.

\bibliographystyle{achemso}
\bibliography{bib}

\end{document}


\def\bibsection{\section*{References}}
\renewcommand{\thepage}{S\arabic{page}}
\renewcommand{\thefigure}{S\arabic{figure}}
\renewcommand{\thesection}{S\arabic{section}}
\renewcommand{\theequation}{S\arabic{equation}}
\title{{Supporting Information}\\ Light-Harvesting Efficiency Cannot Depend on Optical Coherence in the Absence of Orientational Order}

\author{Dominic M. Rouse}
\affiliation{School of Physics and Astronomy, University of Glasgow, Glasgow G12 8QQ, United Kingdom}
\affiliation{Department of Physics and Astronomy, University of Manchester, Manchester M13 9PL, United Kingdom}

\author{Adesh Kushwaha}
\affiliation{School of Chemistry and University of Sydney Nano Institute, University of Sydney NSW 2006, Australia}

\author{Stefano Tomasi}
\affiliation{School of Chemistry and University of Sydney Nano Institute, University of Sydney NSW 2006, Australia}

\author{Brendon W. Lovett}
\affiliation{SUPA, School of Physics and Astronomy, University of St Andrews, St Andrews KY16 9SS, United Kingdom}

\author{Erik M. Gauger}
\affiliation{SUPA, Institute of Photonics and Quantum Sciences, Heriot-Watt University, Edinburgh EH14 4AS, United Kingdom}

\author{Ivan Kassal}
\email[]{ivan.kassal@sydney.edu.au}
\affiliation{School of Chemistry and University of Sydney Nano Institute, University of Sydney NSW 2006, Australia}
\maketitle
\section{Derivation of the time-domain density operator}\label{app:TimeDomain}
In this supporting information, we derive Eq.~(6), closely following ref.~\cite{lavigne2019considerations}. To integrate Eq.~(5), we introduce the interaction picture described by 
\begin{equation}
\begin{aligned}
\mathcal{U}_{SB}(t) &=\exp \left[\mathcal{L}_{SB} t\right], \\
\rho_{I}(t) &=\mathcal{U}_{SB}(-t) \rho(t), \\
\mathcal{V}^{I}_\mu(t) &=\mathcal{U}_{SB}(-t) \mathcal{V}_\mu \mathcal{U}_{SB}(t).
\end{aligned}
\end{equation}
Written in terms of the interaction-picture operators, Eq.~(5) becomes
\begin{equation}\label{eq:der1b}
\frac{\text{d}}{\text{d}t}\rho_I(t)=\sum_{\mu,i}\left( \mb{d}_{\mu}\cdot\mb{e}_i\right)E_i(t)\mathcal{V}^{I}_\mu(t)\rho_I(t)+\text{H.c.}
\end{equation}
Eq.~\eqref{eq:der1b} can be integrated as a power series in $\mathcal{V}^I_\mu(t)$, but since we are interested in weak light, we can truncate the series after the second-order terms. Doing so, switching back to the Schr\"odinger picture, and taking the ensemble average over the stochastic light field yields
\begin{multline}\label{eq:der1c}
    \left\langle\rho(t)\right\rangle=\mathcal{U}_{SB} (t-t_{0}) \rho(t_{0})+ \sum_{i,\mu}\int_{t_{0}}^{t} \mathrm{~d} \tau (\textbf{d}_{\mu}\cdot\textbf{e}_i)\left\langle E_i (\tau)\right\rangle \mathcal{U}_{SB}(t-\tau) \mathcal{V}_{\mu} \mathcal{U}_{SB}(\tau-t_{0}) \rho(t_{0}) + {}\\ \sum_{i,\mu,j,\nu}\int_{t_{0}}^{t} \mathrm{~d} \tau \int_{t_{0}}^{\tau} \mathrm{~d} \tau'  (\textbf{d}_{\mu}\cdot\textbf{e}_i) (\textbf{d}_{\nu}\cdot\textbf{e}_{j})\left\langle E^*_j(\tau)E_i(\tau') \right\rangle\mathcal{U}_{SB}(t-\tau) \mathcal{V}_\mu \mathcal{U}_{SB}(\tau-\tau') \mathcal{V}_\nu \mathcal{U}_{SB}(\tau'-t_{0}) \rho(t_{0}).
\end{multline}
The first order terms in $\mathcal{V}_\mu$ are zero for optical fields because $\left\langle E_i(\tau)\right\rangle=0$. Moreover, assuming that the initial state is stationary, 
\begin{equation}
    \label{eq:assumption}
    \mathcal{U}_{SB}(t) \rho\left(t_{0}\right)=\rho\left(t_{0}\right),
\end{equation}
and defining the retarded Green's functions,
\begin{equation}
    \label{eq:green}
    \mathcal{G}_{SB}(t)=\Theta(t) \mathcal{U}_{SB}(t)=\Theta(t) \exp \left(\mathcal{L}_{SB} t\right),
\end{equation}
we can rewrite Eq.~\eqref{eq:der1c} as
\begin{multline}\label{eq:der2}
    \left\langle\rho(t)\right\rangle=\rho(t_{0})+ \sum_{i,\mu}\int_{t_{0}}^{\infty} \mathrm{~d} \tau (\textbf{d}_{\mu}\cdot\textbf{e}_i)\left\langle E_i (\tau) \right\rangle\mathcal{G}_{SB}(t-\tau) \mathcal{V}_\mu \rho(t_{0}) +\\ \sum_{i,\mu,j,\nu}\int_{t_{0}}^{\infty} \mathrm{~d} \tau \int_{t_{0}}^{\infty} \mathrm{~d} \tau' (\textbf{d}_{\mu}\cdot\textbf{e}_i) (\textbf{d}_{\nu}\cdot\textbf{e}_{j})\left\langle E^*_j(\tau)E_i(\tau')\right\rangle  \mathcal{G}_{SB}(t-\tau) \mathcal{V}_\mu \mathcal{G}_{SB}(\tau-\tau') \mathcal{V}_\nu \rho(t_{0}).
\end{multline}
Introducing the Heaviside $\Theta$ functions in the retarded Green's function enabled us to extend the upper limits of the integrals in Eq.~\eqref{eq:der2} to infinity. This is for later convenience, as it will permit an easier derivation of the Fourier-space density operator in Supporting Information~\ref{app:FrequencySpace}. Finally, Eq.~\eqref{eq:der2} can be written in a more compact notation by collecting terms in the optical coherence matrix $R_{\mu\nu}(\tau,\tau')$ and the propagation matrix $\Lambda_{\mu\nu}(s,s')$, defined in Eqs.~(7) and (10) respectively, resulting in Eq.~(6).

\section{Fourier-space density operator}\label{app:FrequencySpace}
In this supporting information, we rewrite the right-hand side of Eq.~(6) in Fourier space to obtain Eq.~(11). 
Following Ref.~\cite{pachon2013coherent}, we first assume that the field (and therefore $R_{\mu\nu}(\tau,\tau')$) is zero for times before $t_0$, which allows extending the lower limits of integration in Eq.~(6) to $-\infty$, leading to
\begin{equation}\label{eq:rhoApp}
\left\langle\rho(t)\right\rangle=\rho(t_0)+\intf{-\infty}{\infty}{\tau}\intf{-\infty}{\infty}{\tau'}\sum_{\mu,\nu}R_{\mu\nu}(\tau,\tau') \Lambda_{\mu\nu}(t-\tau,\tau-\tau')\rho(t_0).
\end{equation}

To rewrite Eq.~\eqref{eq:rhoApp} in Fourier space, we use the Fourier transforms
\begin{subequations}\label{eq:FTs}
\begin{align}
E_j(t)&=\frac{1}{\sqrt{2\pi}}\intf{-\infty}{\infty}{\omega}\tilde{E}_j(\omega)\rme^{-i\omega t},\\
    \Lambda_{\mu\nu}(s,s')&=\frac{1}{2\pi}\intf{-\infty}{\infty}{\omega_1}\intf{-\infty}{\infty}{\omega_2}\tilde{\Lambda}_{\mu\nu}(\omega_1,\omega_2)\rme^{-i\omega_1 s}\rme^{-i\omega_2 s'}.
\end{align}
\end{subequations}
Substituting Eqs.~\eqref{eq:FTs} into Eq.~\eqref{eq:rhoApp} yields
\begin{equation}\label{eq:derO1}
    \left\langle\rho(t)\right\rangle=\rho(t_0)+\frac{1}{(2\pi)^2}\intf{-\infty}{\infty}{\omega_1}\intf{-\infty}{\infty}{\omega_2}\intf{-\infty}{\infty}{\omega}\intf{-\infty}{\infty}{\omega'}\sum_{\mu,\nu}\tilde{R}_{\mu\nu}(\omega,-\omega')\tilde{\Lambda}_{\mu\nu}(\omega_1,\omega_2)\rme^{-i \omega_1t} I(\omega_1,\omega_2,\omega,\omega')\rho(t_0),
\end{equation}
where $\tilde{R}_{\mu\nu}(\omega,\omega')=\sum_{i,j}\kappa_{\mu\nu ij}\langle\tilde{E}_i(\omega)\tilde{E}_j^*(-\omega')\rangle$ is the Fourier transform of $R_{\mu\nu}(\tau,\tau')$, and
\begin{equation}\label{eq:I}
    I(\omega_1,\omega_2,\omega,\omega')=\intf{-\infty}{\infty}{\tau}\intf{-\infty}{\infty}{\tau'}\rme^{i\tau(-\omega+\omega_1-\omega_2)}\rme^{i\tau'(\omega'+\omega_2)}=\left(2\pi\right)^2\delta(-\omega+\omega_1-\omega_2)\delta(\omega'+\omega_2).
\end{equation}
Substituting Eq.~\eqref{eq:I} into Eq.~\eqref{eq:derO1} yields Eq.~(11).

\section{Controlling the efficiency defined as the ratio of two observables}\label{app:2D}

In this supporting information, we discuss the alternative definition of efficiency mentioned in the text,
\begin{equation}\label{eq:effprime}
    \eta'(t)=\frac{\left\langle\rho_{AA}(t)\right\rangle}{\left\langle\rho_{AA}(t)\right\rangle+\left\langle\rho_{CC}(t)\right\rangle}.
\end{equation}
This definition is a ratio of two simultaneously measurable observables, unlike the definition used in the main text, given in Eq.~(25), which is itself an observable. We show that this difference results in frequency-independent polarization control (FIPC) of $\eta'(t)$ being impossible in some models where FIPC of $\eta(t)$ is possible. We provide a classification system for these models.This discussion holds when comparing any observable to a ratio of two observables for a given system, and not just for the efficiencies $\eta(t)$ and $\eta'(t)$. Therefore, we frame the discussion in this more general sense.

\subsection{General theory}

We consider a system with a set of simultaneously measurable observables $\{O_p(t)\}$. Using Eq.~(11), the expectation value of an observable at time $t$ is 
\begin{equation}\label{eq:Op}
    O_p(t)=O_p(t_0)+\intf{-\infty}{\infty}{\omega_1}\intf{-\infty}{\infty}{\omega_2}\sum_{\mu,\nu}\tilde{R}_{\mu\nu}(\omega_1-\omega_2,\omega_2)\rme^{-i \omega_1 t}\text{Tr}\left(O_p\tilde{\Lambda}_{\mu\nu}(\omega_1,\omega_2)\rho(t_0)\right).
\end{equation}
In the main text, we described how complete orientational and temporal averages can affect $\tilde{R}_{\mu\nu}(\omega,\omega')$ and, in turn, how averages affect observables. For complete averages, if the average rules out control of the observables, then it must also rule out control of a ratio composed of two of the observables. For partial averages, we showed that some control of the observables may still be possible. For most systems, this will mean that control over a ratio of two observables will also be possible. However, we find that this is not always true, because the $\tilde{\mb{n}}(\omega)$ and $\tilde{\phi}(\omega)$ control parameters within a ratio
\begin{equation}\label{eq:r}
    r_{p,p'}(t)=\frac{O_p(t)}{O_{p'}(t)},
\end{equation}
can cancel exactly, making control of $r_{p,p'}(t)$ impossible where control of $\{O_p(t)\}$ is possible. 

From Eq.~\eqref{eq:Op} one can deduce four criteria that must occur simultaneously in order for this cancellation to occur. (1) The averaging cannot be complete so that some control of each observable remains. (2) The initial values of the observables must be zero, $O_p(t_0)=O_{p'}(t_0)=0$, so that the dependence of the second terms in $O_p(t)$ and $O_{p'}(t)$ on $\tilde{\mb{n}}(\omega)$ and $\tilde{\phi}(\omega)$ can cancel out when $O_p(t)$ is divided by $O_{p'}(t)$. This will also mean that the efficiency at $t_0$ is undefined, but this is immediately rectified when $t>t_0$. (3) $\tilde{\mb{n}}(\omega)$ and $\tilde{\phi}(\omega)$ must be frequency-independent so that they can be factored out of the frequency integration in Eq.~\eqref{eq:Op}. Since frequency-independent spectral phase control is just a global field phase it cannot lead to control (i.e., Eq.~\eqref{eq:Op} depends on the spectral phase exclusively through $\phi(\omega)-\phi(\omega')$). Therefore, we will no longer consider temporal averages, instead focusing exclusively on FIPC under partial orientational averages. (4) The polarization control $\tilde{\mb{n}}$ must decouple from the eigenstate state labels $\mu$ and $\nu$ so that one can factor the polarization control out of the summation in Eq.~\eqref{eq:Op}.

We now elaborate on the fourth criterion, since it is the only non-trivial criterion, and defines configurations of transition dipole moments and field properties that allow FIPC of $\eta(t)$ without FIPC of $\eta'(t)$. After enforcing the frequency independence of the control parameters and performing a partial orientational average, denoted by $[\cdot]_{\delta O}$, the optical coherence matrix becomes,
\begin{equation}
    \left[\tilde{R}_{\mu\nu}(\omega,\omega')\right]_{\delta O}=\left\langle\tilde{E}_0(\omega)\tilde{E}_0^*(-\omega')\right\rangle\sum_{i,j}\left[\kappa_{\mu\nu ij}\right]_{\delta O}\tilde{n}_i\tilde{n}_j^*.
\end{equation}
So, the fourth criterion can be stated mathematically as requiring:
\begin{equation} \label{eq:Crit4} \sum_{i,j}\left[\kappa_{\mu\nu ij}\right]_{\delta O}\tilde{n}_i\tilde{n}_j^*=a_{\mu\nu}b(\tilde{\mb{n}}),
\end{equation}
where, importantly, $a_{\mu\nu}$ is independent of $\tilde{\mb{n}}$, and $b(\tilde{\mb{n}})$ is independent of $\mu$ and $\nu$.

Provided that all four of the criteria are satisfied, the value of an observable under partial orientational averaging is 
\begin{equation}
\left[O_p(t)\right]_{\delta O}=b(\tilde{\mb{n}})\intf{-\infty}{\infty}{\omega_1}\intf{-\infty}{\infty}{\omega_2}\sum_{\mu,\nu}a_{\mu\nu}\left\langle \tilde{E}_0(\omega_1-\omega_2)\tilde{E}_0^*(\omega_2)\right\rangle\rme^{-i \omega_1 t}\text{Tr}\left(O_p\tilde{\Lambda}_{\mu\nu}(\omega_1,\omega_2)\rho(t_0)\right),
\end{equation}
and so FIPC of the observable is still possible because the polarization vector $\tilde{\mb{n}}$ appears in the factor $b(\tilde{\mb{n}})$. However, upon taking the ratio of two observables, the factors of $b(\tilde{\mb{n}})$ will cancel, leading to the loss of FIPC. Therefore, the criterion in Eq.~\eqref{eq:Crit4} defines the classification system we set out to derive in this supporting information.

An important question is, for which combination of molecular geometries, field properties, and partial orientational averages is Eq.~\eqref{eq:Crit4} satisfied?

\subsection{Example of a geometry for which $\eta(t)$ is controllable but $\eta'(t)$ is not }

In the main text, we considered an example where a dimer was confined to a two-dimensional plane but free to rotate within that plane. As shown in Fig.~4, we found that polarization control of $\eta(t)$ was possible for both within plane and out of plane dipole moments so long as the light was not propagating perpendicularly to the plane of confinement. In Fig.~4, we repeat this calculation using $\eta'(t)$. One can see that, if the dipole moments are pointing within the plane of confinement, FIPC of $\eta'(t)$ is impossible for all propagation directions of light. 

\begin{figure}[tb]
    \centering
    \includegraphics[width=0.4\textwidth]{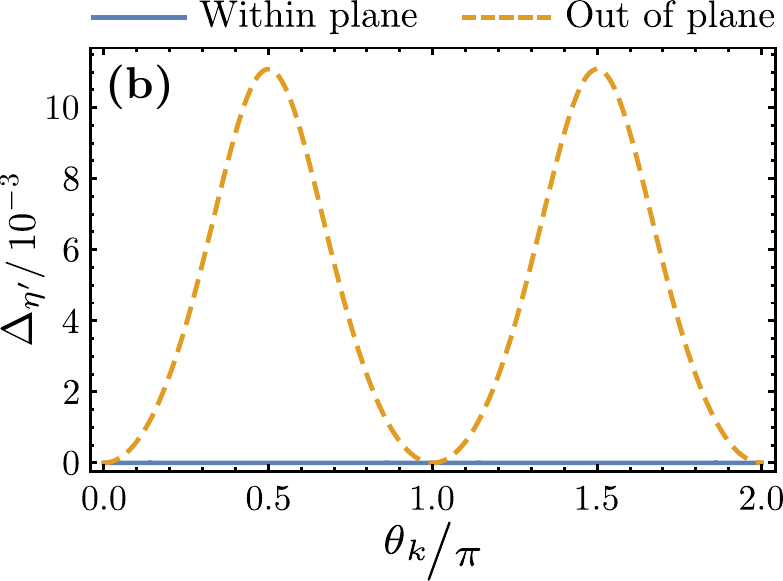}
    \caption{\textbf{Frequency-independent polarization control in two-dimensions using the light-harvesting efficiency $\eta'(t)$.} The model used is the same as in Fig.~4. We plot $\Delta_{\eta'}$, defined as in Eq.~(27) but with $\eta(t_f)$ replaced with $\eta'(t_f)$. Significantly, control of $\eta(t)$ is possible for within plane dipoles if $\theta_k\neq 0,\pi,2\pi$ but is always impossible for $\eta'(t)$. Parameters are given in Fig.~4.}
    \label{fig:pol2dApp}
\end{figure}

We now aim to show that this is predicted by the satisfaction of Eq.~\eqref{eq:Crit4} which requires calculating
\begin{equation}\label{eq:kADO}
  \left[\kappa_{\mu\nu ij}\right]_{\delta O}=\int_{\text{ADO}}  \left(\mb{d}_{\mu}\cdot\mb{e}_i\right) \left(\mb{d}_{\nu}\cdot\mb{e}_j\right) ,
\end{equation}
where `ADO' labels that the integral is over the allowed dipole orientations in the partial average. In Figs.~4 and \ref{fig:pol2dApp}, the dimer has site transition dipole moments with a fixed relative angle of $\xi=\pi/3$ such that  $\mb{d}_L\cdot\mb{d}_R=d_Ld_R\cos(\xi)$. The eigenstate dipole moments also have a fixed relative angle $\tilde{\xi}$ such that $\mb{d}_+\cdot\mb{d}_-=d_+d_-\cos(\tilde{\xi})$, where $\tilde{\xi}$ is a function of the eigenbasis and $\xi$. The dipole vectors are
\begin{equation}
	\mb{d}_\mu(\phi_\mu)=d_\mu\{\sin(\theta_\mu)\cos(\phi_\mu),\sin(\theta_\mu)\sin(\phi_\mu),\cos(\theta_\mu)\},
\end{equation}
for $\mu\in\{+,-\}$ where $\phi_\mu\in[0,2\pi]$ describe the freedom of the molecule to rotate within the $xy$ plane and we have made the dependence on the $\phi_\mu$ explicit. $\theta_\mu$ are fixed and determine whether the dipoles are within plane ($\theta_+=\theta_-=\theta_L=\theta_R=\pi/2$) or out of plane. The angles $\phi_+$ and $\phi_-$ are related through the constant angle $\tilde{\xi}$ by
\begin{equation}\label{eq:zeta}
	\cos(\tilde{\xi})=\cos\left(\phi_+-\phi_-\right)\sin(\theta_+)\sin(\theta_-)+\cos(\theta_+)\cos(\theta_-),
\end{equation}
so that the freedom for the molecule to rotate is described by a single angle, either $\phi_+$ or $\phi_-$ (with the other determined by Eq.~\eqref{eq:zeta}). Therefore, for this geometry, Eq.~\eqref{eq:kADO} is
\begin{equation}
  \left[\kappa_{\mu\nu ij}\right]_{\delta O}=\int_{0}^{2\pi} d\phi_{\mu}\, \left(\mb{d}_{\mu }(\phi_\mu)\cdot\mb{e}_i\right)\left( \mb{d}_{\nu j}(\phi_\nu(\phi_\mu))\cdot\mb{e}_j\right),
\end{equation}
where $\phi_\nu(\phi_\mu)$---that is, $\phi_\nu$ as a function of $\phi_\mu$---is obtained by solving  Eq.~\eqref{eq:zeta}. Performing the integration yields
\begin{equation}\label{eq:k1}
    \left[\kappa_{\mu\nu ii}\right]_{\delta O}=\pi d_\mu d_\nu\begin{cases}
        \delta_{\mu\nu}\sin^2(\theta_\mu)+\left(1-\delta_{\mu\nu}\right)\left(\cos(\tilde{\xi})-\cos(\theta_\mu)\cos(\theta_\nu)\right)&\text{ if }i=x,y\\
        2\delta_{\mu\nu}\cos^2(\theta_\mu)+\left(1-\delta_{\mu\nu}\right)\cos(\theta_\mu)\cos(\theta_\nu)&\text{ if }i=z
        \end{cases}
\end{equation}
and
\begin{equation}\label{eq:k2}
    \left[\kappa_{\mu\nu i,j\neq i}\right]_{\delta O}=
        \pi d_\mu d_\nu\left(1-\delta_{\mu\nu}\right)\begin{cases}
        +F(\tilde{\xi},\theta_\mu,\theta_\nu)&\text{ if }\{i,j\}=\{x,y\}\\
        -F(\tilde{\xi},\theta_\mu,\theta_\nu)&\text{ if }\{i,j\}=\{y,x\}\\
        0&\text{ otherwise},
        \end{cases}
\end{equation}
where $F(\tilde{\xi},\theta_\mu,\theta_\nu)$ is a complicated function which we do not write down because its form is not important beyond what is written in Eq.~\eqref{eq:k2}.

We can now use Eqs.~\eqref{eq:k1} and \eqref{eq:k2} to show that when the criterion in Eq.~\eqref{eq:Crit4} is satisfied, $\eta'(t)$ cannot be controlled whilst $\eta(t)$ can be. To exemplify this, we will consider two directions of light propagation for the dimer restricted to rotations in the $xy$ plane. In the first example, we will choose $z$-directed light, corresponding to $\theta_k\in\{0,\pi,2\pi\}$. From Figs.~4 and \ref{fig:pol2dApp}, we should find that all dependence on $\tilde{\mb{n}}$ vanishes because FIPC of \textit{observables} is impossible. In the second example, we will choose $y$-directed light, corresponding to $\theta_k=\pi/2$, where we should find that Eq.~\eqref{eq:Crit4} is satisfied for within plane dipoles but not for out of plane dipoles.

\textbf{Example 1: $z$-directed light.} For $z$-directed light $i,j\in\{x,y\}$ in the summation on the left-hand-side of Eq.~\eqref{eq:Crit4}. After substituting Eqs.~\eqref{eq:k1} and \eqref{eq:k2} into the left-hand-side Eq.~\eqref{eq:Crit4}, noting that $n_i$ are real in this example, we find that
\begin{equation}\label{eq:Az}
    \sum_{i,j\in\{x,y\}}\left[\kappa_{\mu\nu ij}\right]_{\delta O}\tilde{n}_i\tilde{n}_j^*=\pi d_\mu d_\nu\left(\delta_{\mu\nu}\sin^2(\theta_\mu)+\left(1-\delta_{\mu\nu}\right)\left(\cos(\xi)-\cos(\theta_\mu)\cos(\theta_\nu)\right)\right),
\end{equation}
where we have used $\sum_{i\in\{x,y\}}\tilde{n}_i^2=1$. Since Eq.~\eqref{eq:Az} does not depend on $\tilde{\mb{n}}$, FIPC of observables is impossible regardless of dipole orientations for $z$-directed light. Since FIPC is impossible in the observables, it is also lost in ratios. Interestingly, the loss of FIPC in this case would not occur if $\tilde{\mb{n}}$ were complex, i.e., for circularly or elliptically polarized light. This is because the terms with $i\neq j$, $F(\xi,\theta_\mu,\theta_\nu)(\tilde{n}_x\tilde{n}_y^*-\tilde{n}_x^*\tilde{n}_y)$, only cancel for real $\tilde{n}_i$.

\textbf{Example 2: $y$-directed light.} Substituting Eqs.~\eqref{eq:k1} and \eqref{eq:k2} into the right-hand-side of Eq.~\eqref{eq:Crit4}, this time with $i,j\in\{x,z\}$, yields
\begin{equation}\label{eq:Ay}
    \sum_{i,j\in\{x,z\}}\left[\kappa_{\mu\nu ij}\right]_{\delta O}\tilde{n}_i\tilde{n}_j^*=\pi d_\mu d_\nu\delta_{\mu\nu}\left(\sin^2(\theta_\mu)\tilde{n}_x^2+2\cos^2(\theta_\mu)\tilde{n}_z^2\right)+\left(1-\delta_{\mu\nu}\right)\left(\cos(\xi)\tilde{n}_x^2+\cos(\theta_\mu)\cos(\theta_\nu)\right),
\end{equation}
where we have used $\sum_{i\in\{x,z\}}\tilde{n}_i^2=1$. Eq.~\eqref{eq:Ay} shows that the fourth criterion in Eq.~\eqref{eq:Crit4} is only satisfied if $\cos(\theta_+)=\cos(\theta_-)=0$, i.e. for within plane dipoles, in which case 
\begin{equation}\label{eq:AyWP}
\left.\sum_{i,j\in\{x,z\}}\left[\kappa_{\mu\nu ij}\right]_{\delta O}\tilde{n}_i\tilde{n}_j^*\right\vert_{\text{within plane}}=\pi d_\mu d_\nu\left(\delta_{\mu\nu} +\left(1-\delta_{\mu\nu}\right)\cos(\zeta)\right)n_x^2,
\end{equation}
which is factorised as in Eq.~\eqref{eq:Crit4}. Therefore, within-plane dipoles with $y$-directed light describes a geometry for which FIPC of $\eta'(t)$ is not possible, but FIPC of $\eta(t)$ is, and this is evidenced in Fig.~\ref{fig:pol2dApp}.
\bibliographystyle{achemso}
\bibliography{bib}